\documentclass[twocolumn,showpacs,preprintnumbers,amsmath,amssymb]{revtex4}
\usepackage[dvips]{graphicx}
\usepackage{dcolumn}

\begin{document}

\title
{Superfluid Quantum Criticality in Liquid $^3$He in Anisotropic Aerogel
}
\author{Dai Nakashima and Ryusuke Ikeda}
\affiliation{Department of Physics, Graduate School of Science, Kyoto University, Kyoto 606-8502, Japan}
\date{\today}

\begin{abstract}
In the novel superfluid polar phase realized in liquid $^3$He in highly anisotropic aerogels, a quantum transition to the polar-distorted A (PdA) phase may occur at a low but finite pressure $P_c(0)$. It is shown that a nontrivial quantum dynamics of the critical fluctuation of the PdA order is induced by the presence of both the columnar-like impurity scattering leading to the Anderson's Theorem for the polar phase and the line node of the quasiparticle gap in the state, and that, in contrast to the situation of the normal to the B-phase transition in isotropic aerogels, a weakly divergent behavior of the compressibility appears in the quantum critical region close to $P_c(0)$.
\end{abstract}


\maketitle

\section{Introduction}
It has been found recently that the celebrated Anderson's Theorem \cite{PWA} implying that the bulk properties of an $s$-wave paired superfluid phase are insensitive to the impurity strength is satisfied in the $p$-wave polar superfluid phase \cite{Dmitriev,AI06} of liquid $^3$He realized in highly anisotropic aerogels with a structure consisting of columnar defects \cite{Fomin18,Eltsov,Hisamitsu}. Consequently, the normal to polar superfluid transition there has no quantum critical point (QCP), and the polar phase is stabilized as the high temperature superfluid phase. In contrast, the A and B phases appearing at lower temperatures there, called as the polar-distorted A and B (PdA and PdB) phases \cite{Makinen}, respectively, are not protected by the impurity scattering and are pushed down to lower temperatures in the pressure ($P$) to temperature ($T$) phase diagram \cite{Hisamitsu}. Then, we have a $P$ to $T$ phase diagram in which a QCP of the polar to PdA second order transition line is present at a finite pressure $P_c(0)$.
This situation in which a QCP between the two superfluid phases occurs is exceptional in the context of Fermi superfluids, and it is valuable to see how the present issue is different from the normal to superfluid $B$ QCP realized in liquid $^3$He in isotropic
aerogels \cite{Parpia,London,Thuneberg}.

In the present paper, the quantum fluctuation accompanying the polar to PdA transition is examined both in the fictitious clean limit and in the model of anisotropic aerogels in the limit of columnar defects. It is found that, in the clean limit, the line node of the quasiparticle energy gap in the polar phase makes the dynamics of the non-polar boson (i.e., collective) modes purely dissipative like the fluctuation mode associated with the normal to B transition occurring in isotropic aerogels. In contrast, in the realistic polar phase stabilized by the columnar-like defects where the Anderson's Theorem is satisfied, the corresponding dynamics of the PdA critical fluctuation \cite{com1} in zero temperature limit becomes an oscillating one multiplied by a logarithmic correction. We show that this dynamics is reflected in the compressibility as a divergent behavior like $\sqrt{{\rm log}T^{-1}}$ in the close vicinity of the quantum critical (QC) pressure $P_c(0)$ in contrast to the nondivergent behavior close to the normal to superfluid B QCP occurring in the isotropic aerogels. Therefore, the present result can be regarded as one of the properties characteristic of the novel 3D superfluid polar phase realized in the anisotropic
aerogels.

The present paper is organized as follows. In sec.2, the dynamics of the PdA fluctuation is examined by assuming the polar to PdA transition to occur in clean limit with no impurity scattering. In sec.3, the PdA fluctuation is investigated in the case with columnar impurity scatterings well describing the scatterings due to the real nematic aerogel. In sec.4 and 5, a critical behavior occurring {\it within} the polar superfluid phase due to the QC PdA fluctuation is discussed. In sec.6, the obtained results in the present paper are summarized. Details of the fluctuation-renormalization are explained in Appendix.

\section{Quantum Critical Dynamics of Polar to PdA transition in Clean limit}

As usual, we start from the BCS Hamiltonian for the $p$-wave-paired superfluid occurring in the Fermi liquid with an isotropic Fermi surface. Throughout this manuscript, we work in the weak-coupling approximation, and the strong coupling effect is assumed to have already been included in the microscopic model only as a small effect for stabilizing the PdA phase rather than the PdB phase. Then, the partition function describing the bulk properties and the possible superfluid transitions will be expressed as a functional integral over the order parameter field $A_{\mu,j}$,  $Z=\int {\cal D}A_{\mu,j} {\cal D}A^*_{\mu,j} \exp(-{\cal S})$, where
\begin{equation}
{\cal S} = \frac{1}{3|g| T} \sum_{{\bf q}, \omega} A^*_{\mu,j} A_{\mu,j} - {\rm ln}\biggl\langle T_s \exp {\hat \Pi} \biggr\rangle
\label{S}
\end{equation}
with
\begin{eqnarray}
\Pi &=& \frac{1}{2} \sum_{{\bf q}, \omega} \int_p {\hat p}_j \biggl[ A^*_{\mu,j}({\bf q},\omega) \int_0^{T^{-1}} ds \, e^{{\rm i}\omega s} \, {\hat a}_{p_+,\alpha}(s) \nonumber \\
&\times& {\rm i} (\sigma_2 \sigma_\mu)_{\alpha,\beta} {\hat a}_{-p_-, \beta}(s) + {\rm h.c.} \biggl].
\end{eqnarray}
Here, $|g|$ is the attractive interaction strength, $T$ is the temperature, $\int_p = \int d^3p/(2 \pi)^3$, and ${\hat p}_j = p_j/p_{\rm F}$ ($j=1$, $2$, $3$), $p_{\rm F}/\hbar$ is the Fermi wavenumber, $a_{p,\beta}(s)$ is the fermion operator with the momentum ${\bf p}$, the imaginary time $s$, and the spin index $\beta$. Further, $\langle \cdot \cdot \cdot \rangle$ denotes the ensemble average on the Hamiltonian expressing free fermions. Hereafter, the superfluid order parameter field with the spin and orbital indices $\mu$ and $j$ will be expressed in the form composing of the mean field polar order parameter $\Delta_p=\Delta {\hat p}_3$ and the additional fluctuation $e({\bf r})$ of the PdA state
\begin{equation}
A_{\mu,j}({\bf r}) = \delta_{\mu,2}(\delta_{j,3} \Delta + {\rm i} \delta_{j,1} e({\bf r})).
\label{OP}
\end{equation}
The overall phase of $e({\bf r})$ is taken to be the same as that of the polar order parameter $\Delta$. Consequently, $e({\bf r})$ can be regarded as a real field equivalent to an Ising spin ordering \cite{com2}. Here, a sign-reversal of $e({\bf r})$ corresponds to a reversal of the chirality of the ABM
pairing \cite{Volovik}.

The GL action for the $e$-field can be constructed as usual from ${\cal S}$, and, up to O($e^2$), it takes the form \cite{Ebisawa}
\begin{widetext}
\begin{eqnarray}
{\cal S}^{(2)} &=& {\cal S} - {\cal S}_{\rm polar}(\Delta) = \frac{1}{T} \sum_{{\bf q},\omega} \biggl[\frac{1}{3|g|} |e({\bf q},\omega)|^2 - T\sum_\varepsilon \int_p {\hat p}_1^2 \biggl[ {\cal G}({\bf p}_+,\varepsilon+\omega) {\cal G}(-{\bf p}_-,-\varepsilon) |e({\bf q},\omega)|^2 \nonumber \\
&-& \frac{1}{2} \biggl( {\cal F}({\bf p}_+,\varepsilon+\omega) {\cal F}(-{\bf p}_-,-\varepsilon) e^*({\bf q},\omega) e^*(-{\bf q}, -\omega) + {\cal F}^\dagger({\bf p}_+,\varepsilon+\omega) {\cal F}^\dagger(-{\bf p}_-,-\varepsilon) e({\bf q},\omega) e(-{\bf q}, -\omega) \biggr) \biggr] \biggr],
\label{flucS}
\end{eqnarray}
\end{widetext}
where ${\cal S}_{\rm polar}$ is the mean field free energy in the polar phase divided by $T$. The vertices ${\hat p}_1^2$ arise because the $e$-fluctuation is a non-polar bose excitation in the polar phase.

To examine the dynamics of the PdA fluctuation, for the moment we focus on the terms composed of the Gor'kov Green's functions in eq.(\ref{flucS}). In $T \to 0$ limit, they are expressed by
\begin{widetext}
\begin{equation}
T {\cal S}^{(2)}_{\rm G} = - \sum_{{\bf q},\omega} \int\frac{d\varepsilon}{2 \pi} \int_p {\hat p}_1^2 \frac{(-{\rm i}\varepsilon - \xi_-)({\rm i}(\varepsilon+\omega) - \xi_+) + |\Delta_p|^2}{(\xi^2_- + \varepsilon^2 + |\Delta_p|^2)(\xi^2_+ + (\varepsilon+\omega)^2 + |\Delta_p|^2)} |e({\bf q},\omega)|^2,
\label{flucSG}
\end{equation}
\end{widetext}
where $\xi_\pm = [{\bf p}_\pm^2-p_{\rm F}^2]/(2m)$, ${\bf p}_\pm = {\bf p} \pm {\bf q}/2$, and $\varepsilon$ and $\omega$ are the Fermion and Boson Matsubara frequencies, respectively. To focus on the frequency-dependent terms, just the ${\bf q}=0$ term of eq.(\ref{flucSG}) will be considered in the remainder of this section. Using the Feynman integral $A^{-1}B^{-1} = \int_0^1 d\alpha [ A(1-\alpha)+B\alpha]^{-2}$ and the replacement $\varepsilon \to \varepsilon-\alpha \omega$ \cite{Popov}
and then integrating over $\varepsilon$ and $\xi$, the $q=0$ term of eq.(\ref{flucSG}) becomes
\begin{widetext}
\begin{equation}
T {\cal S}^{(2)}_{\rm G} = \frac{N(0)}{8} \sum_\omega \int_{-1}^1 d{\hat p}_3 (1 - {\hat p}_3^2) \int_0^1 d\alpha \biggl( {\rm ln}\biggl[\frac{(1 - \alpha^2){\overline \omega}^2 + 4 {\hat p}_3^2}{4 x_c}\biggr] + 2 \frac{{\overline \omega}^2(1 - \alpha^2)}{(1 - \alpha^2) {\overline \omega}^2 + 4 {\hat p}_3^2} \biggr) |e(0,\omega)|^2,
\label{flucSpopov}
\end{equation}
\end{widetext}
where ${\overline \omega}=\omega/|\Delta|$, and $x_c$ is the upper cutoff of $(\varepsilon^2 + \xi^2)/\Delta^2$. Further, by performing the $\alpha$-integral, the $\omega$-dependent contribution of ${\cal S}$ is found to become
\begin{widetext}
\begin{eqnarray}
\delta {\cal S}^{(2)}_{\rm G} &=& \sum_\omega \frac{N(0)|{\overline \omega}|}{8T} \int_0^{2/|{\overline \omega}|} d\zeta \, \frac{1 - {\overline \omega}^2 \zeta^2/4}{\sqrt{1+\zeta^2}} \, {\rm ln}\biggl(\frac{\sqrt{1+\zeta^2} + 1}{\sqrt{1+\zeta^2} - 1} \biggr) |e(0,\omega)|^2 \nonumber \\
&=& \frac{\pi^2}{16T} \sum_\omega \frac{N(0)|\omega|}{|\Delta|} \biggl( 1 - \frac{4}{\pi^2} \frac{|\omega|}{|\Delta|} + O\biggl(\frac{\omega^2}{|\Delta|^2} \biggr) \biggr) |e(0,\omega)|^2.
\label{flucSclean}
\end{eqnarray}
\end{widetext}
The $|\omega|$-linear form in eq.(\ref{flucSclean}) indicates that the critical dynamics of the polar to PdA transition in the fictitious clean limit is purely dissipative. This dissipative dynamics is induced by the horizontal line node of the energy gap in the polar phase.

Let us compare this result with those in other familiar situations known so far.  First, in the gapful B phase of the bulk liquid $^3$He, the quantum dynamics of the collective modes is of the $\omega^2$ form \cite{Popov}. However, there is no QCP in the bulk liquid $^3$He.

The present $|\omega|$ term is familiar in the quantum transition between the normal state and a unconventional Fermi superfluid state such as a $d$-wave superconducting phase. In liquid $^3$He in aerogels, such a situation occurs as the normal to $B$-phase 3D transition in isotropic aerogels \cite{Parpia}. In the case, however, this fluctuation dynamics does not lead to any divergent behavior of a second derivative of the free energy which, in the present situation, is the compressibility (see sec.V).

\section{Quantum Critical Dynamics of Polar to PdA Transition in Nematic Aerogel}

Now, let us turn to examining the quantum dynamics of the critical fluctuation of the {\it real} polar to PdA transition in an environment with the elastic scatterings due to the anisotropic aerogel. According to Ref.\cite{Hisamitsu}, we introduce the impurity scattering term
\begin{equation}
{\cal H}_{\rm imp} = \int_{\bf r} \psi^\dagger_\sigma({\bf r}) u({\bf r}) \psi_\sigma({\bf r})
\label{impmodel}
\end{equation}
in the Hamiltonian. Regarding the random averaging over $u({\bf r})$, the Fourier transform $u({\bf k})$ of $u({\bf r})$ is assumed to have zero mean and the mean-squared average
\begin{equation}
{\overline {|u_{\bf k}|^2}} = \frac{1}{2 \pi N(0) \tau} w({\bf k}).
\label{impstructure}
\end{equation}
The strong anisotropy of the scattering events due to the columnar-like aerogel is incorporated in the momentum dependence of $w({\bf k})$. According to the treatment in the standard textbook \cite{AGD}, the impurity-averaged Gor'kov Green's functions are expressed as
\begin{eqnarray}
{\tilde {\cal G}}_{\bf p}(\varepsilon) &=& \frac{-{\rm i}{\tilde \varepsilon}_{\bf p} - \xi}{{\tilde \varepsilon}_{\bf p}^2 + \xi^2 + |{\tilde \Delta}_{\bf p}|^2}, \nonumber \\
{\tilde {\cal F}}^\dagger_{\bf p}(\varepsilon) &=& \frac{ - {\tilde \Delta}_{\bf p}^*}{
{\tilde \varepsilon}_{\bf p}^2 + \xi^2 + |{\tilde \Delta}_{\bf p}|^2},
\label{AGDGF}
\end{eqnarray}
where ${\tilde \varepsilon}_{\bf p}$ and ${\tilde \Delta}_{\bf p}$ satisfy \cite{AGD}
\begin{eqnarray}
i{\tilde \varepsilon}_{\bf p} &=& i\varepsilon - \frac{1}{2 \pi N(0) \tau} \int_{\bf q} w({\bf p} - {\bf q}) {\cal G}_{\bf q}(\varepsilon), \nonumber \\
{\tilde \Delta}_{\bf p} &=& \Delta_{\bf p} - \frac{1}{2 \pi N(0) \tau} \int_{\bf q} w({\bf p} - {\bf q}) [{\cal F}^\dagger_{\bf q}(\varepsilon)]^*,
\label{impVC}
\end{eqnarray}
respectively. One needs to specify the form of $w({\bf k})$ to proceed to detailed calculations. In Ref.\cite{Hisamitsu}, it has been understood that the scattering events in the real nematic aerogels used in experiments \cite{Dmitriev} is well approximated by the limit of the strong anisotropy \cite{Fomin18}. Hence, for simplicity, we set hereafter
\begin{equation}
w_\infty({\bf k}) = \pi k_{\rm F} \delta(k_z).
\label{inftyimpstructure}
\end{equation}
Then, the resulting expressions of the quantities in eq.(\ref{impVC}) take the same form as the corresponding ones in the $s$-wave-paired Fermi superfluid \cite{AGD} and are given by
${\tilde \varepsilon}= E_{\bf p}(\varepsilon) \varepsilon/\sqrt{\varepsilon^2 + |\Delta_p|^2}$, and ${\tilde \Delta}_p = E_{\bf p}(\varepsilon) \Delta_p/\sqrt{\varepsilon^2 + |\Delta_p|^2}$ with \cite{Hisamitsu}
\begin{equation}
E_{\bf p}(\varepsilon)=\sqrt{\varepsilon^2 + |\Delta_p|^2}
+ \frac{\pi}{4 \tau}.
\label{Ep}
\end{equation}

If noting that the pairing vertex ${\hat p}_1$ accompanying the PdA order parameter $e({\bf q},\omega)$, which is within the plane of the line node of the polar energy gap, does not suffer from the impurity vertex correction induced by $w_\infty$, it is straightforward to find that the corresponding expression to eq.(\ref{flucSG}) is given in the present case by
\begin{widetext}
\begin{eqnarray}
{\cal S}^{(2)}_{d,{\rm G}} &=& - \frac{1}{T} \sum_{{\bf q},\omega} T\sum_\varepsilon \int_p {\hat p}_1^2 \biggl[ {\tilde {\cal G}}({\bf p}_+,\varepsilon+\omega) \, {\tilde {\cal G}}(-{\bf p}_-,-\varepsilon) |e({\bf q},\omega)|^2 \nonumber \\
&-& \frac{1}{2} {\tilde {\cal F}}({\bf p}_+,\varepsilon+\omega) \, {\tilde {\cal F}}(-{\bf p}_-,-\varepsilon) e^*({\bf q},\omega) e^*(-{\bf q}, -\omega) - \frac{1}{2} {\tilde {\cal F}}^\dagger({\bf p}_+,\varepsilon+\omega) \, {\tilde {\cal F}}^\dagger(-{\bf p}_-,-\varepsilon) e({\bf q},\omega) e(-{\bf q}, -\omega) \biggr],
\label{flucSdirt}
\end{eqnarray}
\end{widetext}
where ${\tilde {\cal F}}({\bf p},\varepsilon)=({\tilde {\cal F}}^\dagger({\bf p},-\varepsilon))^*$.
Hereafter, the $\omega$ and ${\bf q}$ dependences will be considered separately.  First, we focus on the frequency dependences by setting ${\bf q}=0$. Then, by performing the $\xi$-integral, the ${\bf q}=0$ term of eq.(\ref{flucSdirt}) is expressed at $T=0$ in the form
\begin{equation}
T{\cal S}^{(2)}_{d,{\rm G}}|_{q=0} = N(0) \sum_\omega C_{20}(\omega) |e(0,\omega)|^2,
\end{equation}
where
\begin{widetext}
\begin{equation}
C_{20}(\omega) = - \frac{1}{8} \int_{-1}^1 d{\hat p}_3 (1 - {\hat p}_3^2) \int d\varepsilon \frac{1}{E_{\bf p}(\varepsilon)+E_{\bf p}(\varepsilon+\omega)} \biggl(1 + \frac{\varepsilon (\varepsilon+\omega) + |\Delta_p|^2}{\sqrt{(\varepsilon^2+|\Delta_p|^2)((\varepsilon+\omega)^2+|\Delta_p|^2)}} \biggr).
\label{flucfreqdirt}
\end{equation}
\end{widetext}
To make the ensuing analytical treatment tractable, we focus on the low energy behavior of the above expression which is obtained by replacing the prefactor $1/(E(\varepsilon)+E(\varepsilon+\omega))$ in the integrand by $2 \tau/\pi$ (see eq.(\ref{Ep})). Expecting a nontrivial frequency dependence associated with the gap node to occur and replacing the upper limit of $|{\hat p}_3|$ with infinity, the $\omega$ dependence, $\Delta C_{20}(\omega) \equiv C_{20}(\omega) - C_{20}(0)$, of eq.(\ref{flucfreqdirt}) becomes
\begin{eqnarray}
\Delta C_{20}(\omega) \!\! &\simeq& \!\! \frac{\tau|\Delta|}{4} \int_0^{s_c} ds \biggl( \frac{2}{\pi} \frac{4s - {\overline \omega}^2}{4s+{\overline \omega}^2} K(k) - 1 \biggr) \nonumber \\
&=& \frac{\tau}{8 |\Delta|} \omega^2 \biggl[ {\rm ln}\biggl(\frac{|\Delta|}{|\omega|} \biggr) + {\rm ln}s_c^{1/2} + 1.385 \biggr],
\label{flucfreqdirt1}
\end{eqnarray}
where $k=4 s^{1/2} |{\overline \omega}|/(4s+{\overline \omega}^2)$, $s_c$ is the upper cutoff of ${\hat p}_3^2 + (\varepsilon/\Delta)^2$, and $K(k)$ is the complete elliptic integral of the first kind.
\begin{figure}[b]
\scalebox{1.25}[1.25]{\includegraphics{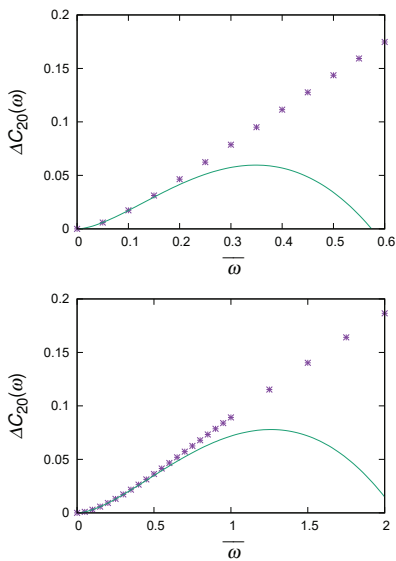}}
\caption{(Color online) (a) Resulting ${\overline \omega}$ ($\equiv |\omega|/|\Delta|$) v.s. $\Delta C_{20}(\omega)$ ($\equiv C_{20}(\omega) - C_{20}(0)$) curves for $\tau |\Delta|=7.85$. Data directly follow from eq.(\ref{flucfreqdirt}), while the solid curve is the corresponding result of eq.(\ref{flucfreqdirt1}). (b) Corresponding results for $\tau |\Delta|=0.785$. We note that, according to Ref.\cite{Hisamitsu}, the experimental phase diagram is explained by using the value of $\tau |\Delta|$ of order unity (see also discussion related to Fig.2). The figures indicate that the frequency range over which eq.(\ref{flucfreqdirt1}) is valid is wider than in the case (a). }
\label{fig.1}
\end{figure}

In Fig.1, the approximated form (\ref{flucfreqdirt1}) of the frequency dependence of the coefficient of $|e(0,\omega)|^2$ in eq.(\ref{flucSdirt}) is compared with the numerical results of the full expression, eq.(\ref{flucfreqdirt}), for the two values of the impurity strength $1/(\tau |\Delta|)$. It is clearly seen that the exact frequency dependence is well approximated by that of the low energy expression eq.(\ref{flucfreqdirt1}) for lower ${\overline \omega}$ values, and that the range of ${\overline \omega}$ in which the low energy expression is valid becomes wider for dirtier cases.

In contrast to the frequency terms, the gradient terms are not significantly affected by the gap nodes and are obtained from
\begin{eqnarray}
\delta {\cal S}^{(2)}_{d,{\rm G}}|_{{\bf \omega}=0} \!\! &=& \!\! \sum_{\bf q} \int\frac{d\varepsilon}{2 \pi T} \int_{\bf p} {\hat p}_1^2 \biggl[ \frac{{\tilde \varepsilon}^2 + \xi^2 + |{\tilde \Delta}_p|^2}{(\xi^2 + E_{\bf p}^2)} \nonumber \\ &-& \frac{({\rm i}{\tilde \varepsilon} - \xi_+)(-{\rm i}{\tilde \varepsilon} - \xi_-)+|{\tilde \Delta}_p|^2}{(\xi^2 + E_{{\bf p}_+}^2)(\xi^2 + E_{{\bf p}_-}^2)} \biggr] |e({\bf q},0)|^2 \nonumber \\
&=& \frac{N(0)}{8T} \sum_{\bf q} \biggl\langle {\hat p}_1^2 \int d\varepsilon \frac{({\bf v}\cdot{\bf q})^2}{E_{\bf p}^3} \biggr\rangle_{\hat p}
|e({\bf q},0)|^2.
\label{flucgraddirt}
\end{eqnarray}
Here, $\langle \cdot \cdot \cdot \rangle_{\hat p}$ denotes the average over the ${\bf p}$-direction.

Although the bulk properties are $\tau$-independent \cite{Fomin18,Hisamitsu} in the present strongly anisotropic limit, the gradient terms
are $\tau$-dependent.
Previously, it has been pointed out that, close to a normal to $p$-wave superfluid transition in an environment with elastic impurity scatterings, the gradient term close to the zero temperature is accompanied by a logarithmically divergent correction \cite{Adachi,Nagamura} as a consequence of the impurity-induced pairing vertex correction. In the present case of the transition to another superfluid phase in the polar superfluid phase, such a vertex correction is absent and hence, no nonanalytic correction arises in the gradient terms.

The resulting gradient terms generally take the form
\begin{equation}
\delta {\cal S}^{(2)}_{d,{\rm G}}|_{{\bf \omega}=0} = \frac{N(0)}{T} \xi_0^2 \sum_{\bf q} c_{ij} q_i q_j |e({\bf q}, 0)|^2,
\end{equation}
where
\begin{equation}
c_{ij} = 3 c_\perp \delta_{ij} + {\hat z}_i {\hat z}_j (c_3 - 3 c_\perp) - 2 {\hat l}_i {\hat l}_j c_\perp.
\label{flucgraddirt1}
\end{equation}
Here, $\xi_0=v_{\rm F}/|\Delta|$. Further, ${\hat z}$ and ${\hat l}$ are the directions of the polar anisotropy and of the gap node in the resulting PdA phase, respectively, and the inequalities $c_3>0$ and $c_\perp > 0$ are satisfied as stability conditions. Note that, in the present case considering the amplitude fluctuation of the order parameter, ${\hat l}$ is assumed to be a constant vector as far as one restricts oneself to the Gaussian fluctuation. A typical value of the impurity strength $1/(\tau |\Delta|)$ in the polar phase in the nematic aerogels has been suggested to be of order unity \cite{Hisamitsu}. For instance, the use of the value $1.10$ of $\tau|\Delta|$ leads to the values of the coefficients $c_3=7.56 \times 10^{-3}$ and $c_\perp = 1.16 \times 10^{-2}$. The smallness of $c_3$ is a reflection of the fact that the polar gap $|\Delta_p|$ is maximal along ${\hat z}$.

\section{Gaussian Fluctuation at zero temperature}

In this section, we will explain how the nonanalytic frequency dependence of the PdA fluctuation in the polar phase should be reflected in a physical quantity at zero temperature. First, the results in the preceding section will be summarized. We have shown that the Gaussian action of the PdA order parameter fluctuation in the polar superfluid phase takes the form
\begin{eqnarray}
T{\cal S}^{(2)}_{d} &=& N(0) \sum_{{\bf q}, \omega} \biggl[ c_m(P) + \frac{|\Delta| \tau}{8} {\overline \omega}^2 (|{\rm ln}|{\overline \omega}|| + c_2) \nonumber \\
&+& \xi_0^2 c_{ij} q_i q_j  \biggr] |e({\bf q}, \omega)|^2,
\label{action2}
\end{eqnarray}
where $c_2={\rm ln}s_c^{1/2}+1.385$ (see eq.(\ref{flucfreqdirt1})), and
\begin{equation}
c_m(P) = \frac{1}{N(0)} \biggl(\frac{1}{3 g} - T \sum_{\varepsilon} \int_{\bf p} \frac{1 - {\hat p}_3^2}{2} \frac{{\tilde \varepsilon}^2 + \xi^2 + |{\tilde \Delta}_p|^2}{(\xi^2 + E^2_{\bf p}(\varepsilon))^2} \biggr).
\label{mass}
\end{equation}
The pressure dependence of the coefficient $c_m$ of the mass term has been evaluated in Ref.\cite{Hisamitsu} in terms of the pressure dependence of the superfluid transition temperature of the bulk liquid $^3$He and takes the form $c_m = {\cal C}(P_c(0) - P)/P_c(0)$ near the QC pressure $P_c(0)$, where the coefficient ${\cal C}$ depend weakly on the impurity strength $\tau |\Delta|$ and is about $0.18$.

The $\tau$-dependence of $P_c(0)$ can be seen through Fig.3 in Ref.\cite{Hisamitsu}. It is found that, in the limit of strong anisotropy, a positive $P_c(0)$ appears in $(2 \pi \tau)^{-1} > 0.45$ (mK), although the PdA phase itself is lost for stronger disorder $(2 \pi \tau)^{-1} > 1.1$ (mK).

The parameter $c_2$ grows with decreasing $|\Delta|\tau$. In contrast, as the impurity scattering is significantly reduced, the sign of $c_2$ becomes negative. This negative $c_2$ is also reflected in the solid curves in Fig.1 and is a vestige of the negative coefficient of the $\omega^2$ term in clean limit (see eq.(\ref{flucSclean})).

As a physical quantity measurable in liquid $^3$He and showing the critical behavior at low temperatures, we study here the compressibility
\begin{equation}
\kappa = - \frac{1}{n^2 V} \frac{\partial^2 F}{\partial \mu^2}
\label{totalkappa}
\end{equation}
where $n$ is the particle density of liquid $^3$He, $V$ is the volume, $\mu$ is the chemical potential, and $F$ is the total free energy. By replacing $\mu$ by the pressure $P$ and using the fact that, near $P_c$, the $P$-dependence of $\kappa$ due to the PdA-fluctuation is dominated by that of the mass term $c_m$ in eq.(\ref{mass}), the resulting fluctuation contribution $\kappa_{\rm fl}$ to $\kappa$ is expressed as
\begin{equation}
\kappa_{\rm fl} = T \biggl(\frac{{\cal C}}{n \, P_c} \frac{\partial P}{\partial \mu}\biggr|_{P_c} \biggr)^2 \sum_{\omega} \int_{\bf q} \frac{1}{[ c_m + \Delta C_{20}(\omega) + \xi_0^2 c_{ij} q_i q_j ]^2}
\label{fluccomp}
\end{equation}
At zero temperature, this expression can be rewritten as
\begin{equation}
\kappa_{\rm fl} = \biggl(\frac{{\cal C}}{n P_c} \frac{\partial P}{\partial \mu}\biggr|_{P_c} \biggr)^2 \frac{|\Delta|}{2 \pi c_\perp \xi_0^3} \sqrt{\frac{2}{3 c_3 \tau |\Delta|}} \,\, I( x ; c_2)
\label{fluccomp1}
\end{equation}
where
\begin{equation}
I(x; c_2) = \int_0^{{\overline \omega}_c} \frac{d{\overline \omega}}{2 \pi} \frac{1}{\sqrt{x + {\overline \omega}^2 ({\rm ln}{\overline \omega}^{-1} + c_2)}},
\label{integ}
\end{equation}
and
\begin{equation}
x = \frac{8 {\cal C}}{\tau |\Delta|} \biggl(1 - \frac{P}{P_c} \biggr).
\label{x}
\end{equation}


\begin{figure}[t]
\scalebox{0.35}[0.35]{\includegraphics{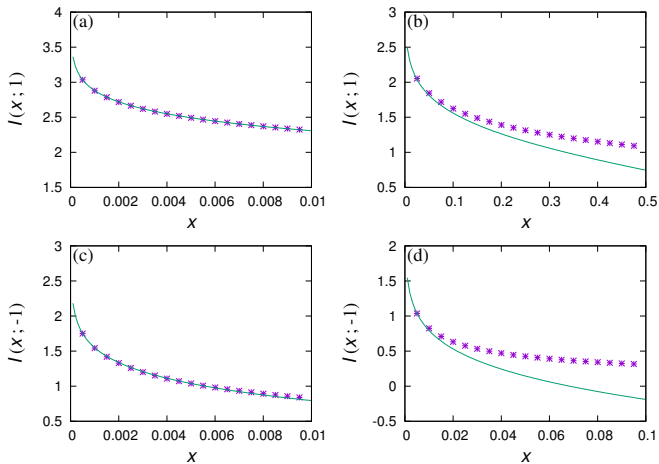}}
\caption{(Color online) Computed $I(x;c_2)$-curves for a moderately impure case with $c_2=1.0$ (upper two figures (a) and (b)) and for a less impure case with $c_2=-1.0$ (lower two figures (c) and (d)). Each solid curve is the $\sqrt{|{\rm log}|x||}$-curve best-fitted to the numerical data (symbols) of $I(x;c_2)$ for each case. To improve the convergence, the frequency cutoff ${\overline \omega}_c$ has been changed for the two cases and was chosen to be $1.0$ for (a) and (b) and to be $0.1$ for (c) and (d). In both cases, at low enough $x$, the $\sqrt{|{\rm log}|x||}$-behavior is well satisfied.
 }
\label{fig.2}
\end{figure}

\begin{figure}[t]
\scalebox{0.75}[0.75]{\includegraphics{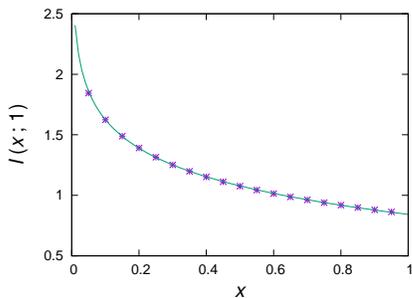}}
\caption{(Color online) The curve of the computed $I(x;1.0)$ in Fig.2 extended over larger $x$-values. The data (symbols) nicely obeys the $|{\rm log}|x||$-form (the solid curve), implying that the ${\rm log}|\omega|$-correction in $\Delta C_{20}$ is ineffective, so that the ordinary $z=1$ 3D GL action is valid, in the range $0.1 < x < 1.0$.
 }
\label{fig.3}
\end{figure}

Examples of the numerically computed $I(x)$ are plotted in Fig.2 for the two values of $c_2$. Our knowledge on a realistic $c_2$-value is based on our previous study on the $P$-$T$ phase diagram \cite{Hisamitsu}. The impurity strength $1/(2 \pi \tau)=0.7$ (mK) used in Ref.\cite{Hisamitsu} (see Fig.3 there) to qualitatively explain the experimental phase diagram in the nematic aerogels \cite{Dmitriev} corresponds to the value $1.1$ of the normalized impurity strength $\tau|\Delta|$, or equivalently to the value $c_2=0.6$.
Through the form of the $\omega$-integral in $I(x)$, it is generally expected that, {\it far from} the QCP at which $x=0$, $I(x)$ behaves like ${\rm log}|x|^{-1}$ which is nothing but the corresponding behavior of the $z=1$ 3D GL model at zero temperature, where $z$ is the dynamical critical exponent. In other words, the ${\rm log}$-correction in $\Delta C_{20}(\omega)$ is ineffective far from the $x=0$ point. On the other hand, the ${\rm log}|\omega|$-correction will change the behavior of $I(x)$ to the $\sqrt{|{\rm log}|x||}$-form closer to
$x=0$. The results presented in Fig.2 and Fig.3 justify this expectation.

We note that the above-mentioned value $c_2=0.6$ is in the intermediate range of the two-values in Fig.2, and that the coefficient $8{\cal C}/(\tau |\Delta|)$ is of order unity when $c_2=0.6$.
By combining this fact with the curves in Fig.2, one can expect the $\sqrt{|{\rm log}(P_c-P)|}$-behavior close to $P_c$ and the crossover to the $|{\rm log}(P_c-P)|$-behavior at lower $P$ of the compressibility $\kappa$ to be visible experimentally as a fluctuation-induced enhancement of $\kappa_{\rm fl}$ at low enough temperatures and at slightly lower pressures than $P_c$.

\section{Quantum Critical Region at Finite Temperatures}

As indicated in the preceding section, the nature of the superfluid fluctuation occurring in the polar phase is closer to that of the $z=1$ 3D GL type rather than the $z=2$ 3D GL one in the case of the purely dissipative dynamics discussed in sec.2. Thus, the QCP of the polar to PdA transition in the nematic aerogel should show a stronger fluctuation effect than that of the normal to superfluid B transition in isotropic aerogels. On the other hand, the $z=1$ 3D system at zero temperature belongs to that at the upper critical dimension so that the genuine critical width of the pressure is negligibly narrow. Nevertheless, at nonzero temperatures, the quantum fluctuation behavior is visible in the so-called quantum critical region, i.e., in the close vicinity of $P=P_c(0)$, reflecting the temperature dependence of the correlation length $\xi_{\rm cr}$ \cite{Sachdev}. Below, we show that, in the present case of the polar to PdA transition, the temperature dependence of $\xi_{\rm cr}$ at low enough $T$ is determined by the mode-coupling, i.e., the interaction between the fluctuations.

To do this, the mode-coupling term
\begin{eqnarray}
T {\cal S}_{\rm int} &=& N(0) u \int_{\bf r} \sum_{\omega_1, \omega_2, \omega_3} e_{-\omega_1}({\bf r}) e_{-\omega_2}({\bf r}) \nonumber \\
&\times& e_{\omega_3}({\bf r}) e_{\omega_1+\omega_2-\omega_3}({\bf r}),
\label{intaction}
\end{eqnarray}
will be added to the Gaussian term, eq.(\ref{action2}). The coefficient $u$ is positive because the polar to PdA transition is of second order as is already known through the previous studies \cite{AI06}. By treating ${\cal S}_{\rm int}$ in the Hartree approximation, the resulting selfenergy correction $\Sigma$ will be incorporated in the fluctuation propagator ${\cal D}({\bf q}, \omega)$ in the form of the renormalization of $c_m$. That is, ${\cal D}({\bf q}, \omega)$ is given by $1/[2(c_m^{(R)} + \Delta C_{20}(\omega) + \xi_0^2 c_{ij} q_i q_j )]$, where the renormalized mass $c_m^{(R)}$ is defined through
\begin{eqnarray}
c_m^{(R)} &=& c_m + \Sigma, \nonumber \\
\Sigma &=& {\overline u} T \sum_\omega \int_{\bf q} {\cal D}({\bf q}, \omega),
\label{renmass}
\end{eqnarray}
where ${\overline u}=u/N(0)$.
\begin{figure}[t]
\scalebox{0.7}[0.7]{\includegraphics{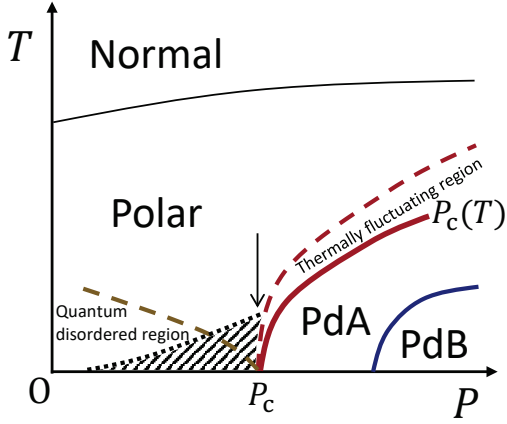}}
\caption{(Color online)
Schematic phase diagram near the QC point $P_c(0)$ of the polar to PdA second order transition curve $P_c(T)$ (thick red solid curve). The thin solid curve and the blue solid curve denote the normal to polar second order transition line and the PdA to PdB first order transition line, respectively. The dashed curves imply the crossover lines on entering the thermal fluctuation region in $P > P_c(T)$ and the quantum disordered region in $P < P_c(0)$, which is typically expressed as $1 - P/P_c(0) \sim 1.6 \times 10^{-2}(T/T_{c0})^2$. The divergent quantum critical behavior of $\kappa$ is expected to be seen on cooling along the downward arrow. The behavior eq.(\ref{estimate}) or $\kappa \simeq \sqrt{|{\rm log}(P_c-P)|}$ one should be seen in the hatched region.
 }
\label{fig.4}
\end{figure}

Before examining $c_m^{(R)}$ in details, one needs to know the $T$-dependence of the bare mass $c_m$, which can be found in terms of the standard technique on the analytic continuation \cite{Ebisawa,AGD}. By replacing the first term in eq.(\ref{mass}) by $\pi T \sum_{\varepsilon}|\varepsilon|^{-1}$ at $T=T_{c0}(P)$, where $T_{c0}(P)$ is the superfluid transition temperature of the bulk liquid, $c_m$ is expressed by $c_m^{({\rm MF})}(0) + \delta c_m(T)$, where
\begin{eqnarray}
c_m^{({\rm MF})}(0) &=& \frac{1}{3} \int_0^{\Lambda} \frac{d\varepsilon}{\varepsilon} {\rm tanh}\biggl(\frac{\varepsilon}{2 T_{c0}} \biggr) \nonumber \\
&-& \int_0^1 d{\hat p}_3 \frac{1 - {\hat p}_3^2}{2} \int_{|\Delta_p|}^{\Lambda} d\varepsilon \frac{\sqrt{\varepsilon^2 - |\Delta_p|^2}}{\varepsilon^2 - |\Delta_p|^2 + (\pi/4 \tau)^2}, \nonumber \\
\delta c_m(T) &=& \int_0^1 d{\hat p}_3 \frac{1 - {\hat p}_3^2}{2} \int_{|\Delta_p|}^{\infty} d\varepsilon \frac{\sqrt{\varepsilon^2 - |\Delta_p|^2}}{\varepsilon^2 - |\Delta_p|^2 + (\pi/4 \tau)^2} \nonumber \\
&\times& (\exp(\varepsilon/T)+1)^{-1}.
\label{MFmass}
\end{eqnarray}
We have numerically checked that the $T=0$ term $c_m^{({\rm MF})}(0)$ is independent of the upper energy cutoff $\Lambda$ and takes the value ${\rm ln}(|\Delta|/T_{c0})^{1/3} + 0.0744$ when $\pi/(4 \tau |\Delta|)=1.0$. On the other hand, $\delta c_m$ is determined by the contribution from the polar gap nodes at lower temperatures and, by power counting, is found to be proportional to $T^3$.

Examining the mode-coupling term $\Sigma$ can be also straightforwardly performed using the analytic continuation for the fluctuation propagator ${\cal D}({\bf q}, \omega)$. Details of its derivation will be given in Appendix. It is found that, at finite but low enough temperatures, the renormalized mass $c_m^{(R)}$ at $P=P_c(0)$ is given by
\begin{widetext}
\begin{equation}
c_m^{(R)} = \delta c_m(T) + 16 \frac{{\overline u}}{|\Delta| \tau} \, \biggl(\frac{T^2}{|\Delta|} \biggr) \biggl( {\rm ln}\biggl(\frac{|\Delta|}{T} \biggr) \biggr)^{1/2} \int_{\bf q} |{\tilde q}|^{-1} f_{\rm B}(|{\tilde q}|),
\label{Rmass}
\end{equation}
\end{widetext}
where $f_{\rm B}(x) = 1/(\exp(x) - 1)$ is the Bose distribution function, and ${\tilde q}^2 = 8 \xi_0^2 c_{ij} q_iq_j/(|\Delta| \tau)$. Since the mean field term $\delta c_m \propto T^3$ is negligible at lower temperatures, the $T$-dependence of $c_m^{(R)}$ at $P=P_c(0)$ and thus, of $P_c(T)$ close to zero temperature is dominated by the mode-coupling term proportional to
$T^2 |{\rm log}T|^{1/2}$.
Therefore, along the $P=P_c(0)$-line at nonzero temperatures, we have
\begin{equation}
\xi_{\rm cr}(T) \simeq \frac{\xi_0}{\sqrt{c_m^{(R)}(P=P_c(0))}} \simeq (T |{\rm log}T|^{1/4})^{-1}.
\label{xiT}
\end{equation}

By combining eq.(\ref{xiT}) with the content in the final two paragraphs in sec.IV, we expect that, at low enough temperatures under a fixed pressure in the close vicinity of $P_c(0)$, $\kappa_{\rm fl}$ behaves like
\begin{equation}
\frac{\kappa_{\rm fl}}{\kappa_n} \sim 10^{-4} (1 - 0.6(T/T_{c0})^3) |{\rm log}(T/T_{c0})|^{1/2},
\label{estimate}
\end{equation}
where we have replaced $\kappa-\kappa_{\rm fl}$ by the compressibility $\kappa_n$ in the normal phase by using the fact that, in the BCS Fermi superfluid, the particle-hole symmetry is well satisfied so that $\partial T_{c0}/\partial \mu$ is extremely small. In eq.(\ref{estimate}), the sub-dominant ${\rm log}|{\rm log}T^{-1}|$ correction has been neglected. It should be noted that the behavior (\ref{estimate}) also arises from $\delta c_m(T)$ without the mode-coupling term included, so that it is not easy to identify the contributions of the mode-coupling term through $\kappa$.

In deriving eq.(\ref{estimate}), we have used eqs.(\ref{totalkappa}) and (\ref{fluccomp1}) and the typical values of $c_\perp$, $c_3$, and ${\cal C}$ mentioned in sec.III and IV together with the $T^3$ behavior of $|\Delta(T)|$ \cite{Hisamitsu}. Further, we have used $k_{\rm F} \xi_0 = 10^2$ and $|\Delta(0)|\tau=1.0$ as their typical values.

A schematic phase diagram to be expected from our analysis is drawn in Fig.\ref{fig.4}. Equation (\ref{estimate}) is valid in the hatched region in Fig.4, and, at higher temperatures, its $|{\rm log}(T/T_{c0})|^{1/2}$-factor is replaced by $|{\rm log}(T/T_{c0})|$ according to the results seen in Figs.2 and 3.
Although eq.(\ref{estimate}) is small in magnitude due to its numerical factor, we expect a deviation due to the logarithmic increase from the saturated $T^3$ behavior of $|\Delta(T)|/|\Delta(0)|$ upon cooling to be accessible in experiments.

These features appearing close to the QC polar to PdA transition are in contrast to the nondivergent behavior of the compressibility near the normal to B QC pressure $P_c^{({\rm NB})}$ in the isotropic aerogel. In fact, it can be seen by replacing $\Delta C_{20}$ in eq.(\ref{fluccomp}) with $|\omega|$ that, along $P=P_c^{({\rm NB})}$ and upon cooling in the case, $\kappa_{\rm fl}$ results in a $T$-independent positive constant $- $ O($T^{1/2}$).

\section{Conclusion}

In the present paper, we have focused on the quantum critical (QC) behavior accompanying the second order transition between the polar and polar-distorted A (PdA) superfluid phases at zero temperature, which can occur in superfluid $^3$He in nematic aerogels, as one of the properties characterizing the novel superfluid polar phase \cite{Dmitriev,AI06}. It has been stressed that the QC behavior of this transition occurring in the anisotropic aerogel cannot be described correctly based on the model in clean limit with no impurity scattering effects. Further, in contrast to the familiar normal to a superfluid quantum transition with the purely diffusive critical dynamics in the isotropic aerogel, the present quantum transition has the dynamics close to that of the $z=1$ 3D GL model, and we can expect the resulting divergent quantum critical behavior of the compressibility in the temperatures and pressure to be experimentally accessible.

\section{Appendix}

Here, our derivation of eq.(\ref{Rmass}) will be explained. Using the spectral representation of the Green's function ${\cal D}({\bf q}, \Omega)$, $\Sigma$ in eq.(\ref{renmass}) is rewritten as
\begin{equation}
\Sigma = {\overline u} |\Delta| \int_{\bf q} \int_0 \frac{d\Omega}{\pi} \, {\rm coth}\biggl(\frac{\Omega |\Delta|}{2T}\biggr) {\rm Im}D_R({\bf q}, -{\rm i}\Omega+\delta),
\end{equation}
where $\delta$ is an infinitesimal positive constant, and $D_R$
is the retarted Green's function corresponding to ${\cal D}$ and takes the form $[D_R({\bf q}, -{\rm i}\Omega + \delta)]^{-1} = c_m^{(R)} + \xi_0^2 c_{ij}q_iq_j - \gamma (\Omega+{\rm i}\delta)^2 [{\rm ln}|\Omega|^{-1} + {\rm i} {\rm sgn}(\Omega) \, \pi/2]$, where $\gamma=|\Delta|\tau/8$. The zero temperature contribution
\begin{equation}
\Sigma(0) = {\overline u} |\Delta| \int_{\bf q} \int_0 \frac{d\Omega}{\pi} \, {\rm Im}D_R({\bf q}, -{\rm i}\Omega+\delta),
\end{equation}
which is positive, will be combined with the corresponding mean field term $c_m^{({\rm MF})}(0)$ (see eq.(\ref{MFmass})). Then, the QCP pressure $P_c(0)$ is defined by $c_m^{({\rm MF})}(0) + \Sigma(0)=0$. To see the $T$ dependence of
\begin{equation}
\Sigma - \Sigma(0) = 2 {\overline u} |\Delta| \int_{\bf q} \int_0 \frac{d\Omega}{\pi} \, f_{\rm B}(y) \, {\rm Im}D_R({\bf q}, -{\rm i}\Omega + \delta),
\end{equation}
where $y=\Omega |\Delta|/T$. We note that, at low enough temperatures, $[T^2 D_R({\bf q}, -{\rm i}\Omega + \delta)]^{-1}$ is approximated by $T^{-2}(c_m^{(R)} + \xi_0^2 c_{ij}q_iq_j) - (y+{\rm i}\delta)^2 |{\rm ln}(T/|\Delta|)|$. Then, by integrating over $\Omega$, one finds that, at $P=P_c(0)$, $c_m^{(R)}$ is given by the expression (\ref{Rmass}).


\begin{thebibliography}{99}
\bibitem{PWA} P. W. Anderson, J. Phys. Chem. Sol. {\bf 11}, 26 (1959).
\bibitem{Dmitriev} V.V. Dmitriev, A.A.Senin, A.A.Soldatov, and A.N.Yudin,et al., Phys. Rev. Lett. {\bf 115}, 165304 (2015).
\bibitem{AI06} K. Aoyama and R. Ikeda, Phys. Rev. B {\bf 73}, 060504(R)
(2006).
\bibitem{Fomin18} I. A. Fomin, JETP {\bf 127}, 933 (2018).
\bibitem{Eltsov} V. B. Eltsov, T. Kamppinen, J. Rysti, and G. E. Volovik, arXiv: 1908.01645.
\bibitem{Hisamitsu} T. Hisamitsu, M. Tange, and R. Ikeda, Phys. Rev. B {\bf 101}, 100502(R)(2020).
\bibitem{Makinen} J. T. Makinen, V. V. Dmitriev, J. Nissinen, J. Rysti, G. E. Volovik, A. N. Yudin, K. Zhang, and V. B. Eltsov, Nat. Comm. {\bf 10},
237 (2019).
\bibitem{Parpia} K. Matsumoto, J. V. Porto, L. Pollack, E. N. Smith, T. L. Ho, and J. M. Parpia, Phys. Rev. Lett. {\bf 79}, 253 (1997).
\bibitem{London} W. P. Halperin, J. M. Parpia, and J. A. Sauls, Phys.Today {\bf 71}, 11, 31 (2018).
\bibitem{Thuneberg} E. V. Thuneberg, S. K. Yip, M. Fogelstrom, and J. A. Sauls, Phys. Rev. Lett. {\bf 80}, 2861 (1998).
\bibitem{com1} The PdA fluctuation on which we focus in this paper is Gaussian in nature. Nevertheless, the fluctuation-renormalization due to the mode-coupling needs to be taken into account only for a possible shift of the transition line (see sec.V).
\bibitem{com2} A similar treatment neglecting the massive relative phase mode [A. J. Leggett, Prog. Theor. Phys. {\bf 36}, 901 (1966)] has been used in a different context. See Ref.10 in [L. Radzihovsky, J. Park, and P. Weichman, Phys. Rev. Lett. {\bf 92}, 160402 (2004)].
\bibitem{Volovik} G. E. Volovik, "The Universe in a Helium Droplet" (Oxford University Press, 2013).
\bibitem{Ebisawa} See, for instance, K. Maki and H. Ebisawa, Prog. Theor. Phys. {\bf 50}, 1452 (1973).
\bibitem{Popov} V. N. Popov, "Functional Integrals in Quantum Field Theory and Statistical Physics" (D.Reidel Publishing Company, 1983).
\bibitem{AGD} A. A. Abrikosov, L. P. Gor'kov, and I. E. Dzyaloshinski, "Methods of Quantum Field Theory in Statistical Physics" (Dover, 1961).
\bibitem{Adachi} H. Adachi and R. Ikeda, J. Phys. Soc. Jpn. {\bf 70}, 2848 (2001).
\bibitem{Nagamura} N. Nagamura and R. Ikeda, arXiv: 1905.02569.
\bibitem{Millis} A. J. Millis, Phys. Rev. B {\bf 48}, 7183 (1993); R. Ramazashvili and P. Coleman, Phys. Rev. Lett. {\bf 79}, 3752 (1997).
\bibitem{Sachdev} S. Sachdev, "Quantum Phase Transitions (second edition)" (Cambridge University Press, 2011) Chapter 14.

\end{thebibliography}
\end{document}